\documentclass[prl,showpacs,twocolumn]{revtex4}
\usepackage{graphicx}
\usepackage{amsmath}
\usepackage{amssymb}
\usepackage{bm}

\begin{document}

\title{Adiabatic pumping in the mixed-valence and Kondo regimes}

\author{Tomosuke Aono}
\affiliation{%
Department of Physics,
Ben-Gurion University of the Negev,
Beer-Sheva 84105, Israel}
\date{\today}
%

\begin{abstract}
We investigate adiabatic pumping through
a quantum dot with a single level in the mixed-valence and Kondo regimes
using the slave-boson mean field approximation.
The pumped current is driven by a gauge potential due to
time-dependent tunneling barriers as well as by the modulation of
the Friedel phase.
The sign of 
the former contribution
depends on the strength of the Coulomb interaction.
Under finite magnetic fields,
the separation of
the spin and charge currents
peculiar to
the Kondo effect
occurs.
\end{abstract} 

\maketitle

{\it Introduction.---}
Adiabatic pumping in a quantum dot system occurs when
the dot is
under slowly varying external gate voltages 
with zero bias voltage.
After a certain period when
the system returns to its initial state,
a finite charge is transferred by
electron interference through the system.
In recent experiments,
both charge~\cite{Switkes99} and spin pumping~\cite{Watson03} are
realized in open quantum dot systems.
This pumping has been investigated theoretically;
charge pumping
~\cite{Brouwer98,Zhou99,Levinson00,Shutenko00,Wei00,Vavilov01,Entin02_2,Moskalets02,Kashcheyevs03}
as well as spin pumping
~\cite{Sharma01,Mucciolo02,Aono03,Governale03,Wang03,Zhou03,Sharma03}.
It has been elucidated that the pumping can be understood in terms of  the Berry phase argument in Ref.~\cite{Zhou03} and Refs.~\cite{Andreev00,Avron00,Makhlin01,DCohen02}.

Adiabatic pumping is investigated for
the systems under  electron-electron interactions
~\cite{Aleiner98,Andreev01,Blaauboer01,Sharma01,Wang02,Aono03}.
In quantum dots,
the interactions introduce a prominent feature,
the Kondo effect
~\cite{Glazman88,Ng88,Goldhaber98,Kouwenhoven01},
where the spin-exchange between electrons in the leads and the dot
results in an enhancement of the conductance at low temperatures.
We investigate the adiabatic pumping in the mixed-valence and the Kondo regimes
to demonstrate an interplay between the Kondo effect and
a gauge potential associated with
the Berry phase,
though this system itself is partly studied in Ref.~\cite{Wang02}.
To this end,
we will show explicitly
the appearance of the Berry phase term due to
time-dependent tunneling barriers  and 
elucidate the connection with the adiabatic pumping
under the electron correlations.
We will then show
that
the spin-charge separation peculiar to the Kondo effect
emerges as
the separation of pumped spin and charge
under a finite magnetic field.

{\it Model.---}
We consider a system which consists a quantum dot
which has a single energy level and couples
to two leads,
described by
the Anderson model~\cite{Glazman88,Ng88}:
$
H =
\sum_{k,\sigma \alpha= L,R} 
 [\epsilon_{k} - \mu(t)] c^{\dagger}_{k\sigma\alpha} c_{k\sigma\alpha}
+ \sum_{k,\sigma \alpha=L,R}
V_{\alpha}(t) \left( c^{\dagger}_{k\sigma\alpha}
d_{\sigma} +
\textrm{h.c.} 
\right)+H_{\rm dot}
$
with $H_{\rm dot} = \sum_{\sigma=\pm}
  E_0 d^{\dagger}_{\sigma} d_{\sigma} +
u\; n_{+} n_{-}$.
Here
$c^{\dag}_{k \sigma \alpha}$ creates an electron
with
energy $\epsilon_{k}$ and spin $\sigma$
in lead  $\alpha = L, R$,
$d^{\dag}_{\sigma}$ creates an electron in
the dot with spin $\sigma$,
$n_{\sigma} = d^{\dagger}_{\sigma} d_{\sigma}$, and
$u$ is the strength of the Coulomb interaction.
We have introduced  time-dependent sources,
a common chemical potential $\mu(t)$ and 
tunneling barriers $V_{\alpha}(t)$,
which are driving forces for electron pumping.

Let us first find out a source-drain voltage
embedded  in the model.
To this end,
we apply the following unitary transformation for
electrons in the leads:
\begin{eqnarray}
\begin{pmatrix}
 c_{k \sigma +}\\
 c_{k\sigma -}
\end{pmatrix}
  =
\begin{pmatrix}
            \xi & \eta\\
            -\eta & \xi
\end{pmatrix}
\begin{pmatrix}
 c_{k \sigma L}\\
 c_{k \sigma R}
\end{pmatrix}
\equiv U^{-1}
\begin{pmatrix}
 c_{k \sigma L}\\
 c_{k \sigma R}
\end{pmatrix}
\end{eqnarray}
with  $\xi=V_{L}/V$, $\eta=V_{R}/V$ and $V = \sqrt{|V_{L}|^2 +
|V_{R}|^2}$~\cite{Glazman88}.
If $V_{\alpha}$ are constant
the Hamiltonian 
reduces
to the single impurity Anderson model;
The modes $c_{k \sigma -}$ decouple with electrons in the dot while
the modes $c_{k \sigma +}$ couple with them
through the tunneling matrix element $V$.

When $V_{\alpha}$ are time-dependent,
the unitary transformation generates
the Berry phase term,
 $(c^{\dagger}_{k \sigma +},  c^{\dagger}_{k \sigma -}) 
 [ - i \hbar U^{-1}(t) \partial U(t)/\partial t ]
(c_{k \sigma +},  c_{k \sigma -})^{t}$=
- $\sum_{k \sigma}
 \left(
c^{\dagger}_{k \sigma +}
 i \hbar a(t)
c_{k \sigma -} -
c^{\dagger}_{k \sigma -}
i \hbar a(t)
c_{k \sigma +}
\right)
$
with a gauge potential $a(t) = (-\xi \dot{\eta} + \dot{\xi} \eta)$;
the modes $c_{k \sigma +/-}$ are mixed.
To diagonalize the  Hamiltonian again,
we apply an additional time-{\it independent} unitary transformation,
$b_{k \sigma s} = 1/\sqrt{2} ( c_{k \sigma +} + i s  c_{k \sigma -} )$
($s=\pm$),
resulting in
\begin{eqnarray}\label{eq:Hamiltoninan_Berry}
H &=&
\sum_{k,\sigma, s=\pm} 
b^{\dagger}_{k \sigma s} 
\left[ \epsilon_{k} -( \mu(t)+ s \hbar a(t) )\right]
b_{k \sigma s} \nonumber\\
&&
+ \frac{V(t)}{\sqrt{2}} \sum_{k,\sigma, s=\pm}
\left[ b_{k \sigma s}^{\dag} d_{\sigma}
+\textrm{h.c.} \right] +
H_{\rm dot}.
\end{eqnarray}
Equation (\ref{eq:Hamiltoninan_Berry}) defines that
the gauge potential $a(t)$,
which originates from
the time-dependent barriers,
acts as a finite bias voltage
between $b_{k \sigma +}$ and $b_{k \sigma -}$ fields.
Hence a finite current flows through the dot
even when a  real bias voltage is zero.
In the following,
we assume $u=0$ for a while.

{\it Adiabatic Current and pumped charge.---}
The current
$I_{\sigma} = \langle \hat{I}_{\sigma, L}-\hat{I}_{\sigma, R} \rangle /2 $
with
$\hat{I}_{\sigma; \alpha}= i e/\hbar\; \sum_{k}
V_{\alpha}(t)
\left(
d^{\dagger}_{\sigma} c_{k \sigma \alpha}
-
c_{k \sigma \alpha}^{\dagger} d_{\sigma} 
\right)
$
can be represented by $b^{\dagger}_{k \sigma s}$ fields:
 \begin{equation}\label{eq:current_definition}
I_{\sigma} =  \frac{ i e V(t)}{2 \sqrt{2}\hbar } 
 \sum_{k \atop s=\pm}
[
 (\xi + i s\eta )^2 \; \langle d^{\dag}_{\sigma} b_{k\sigma s} \rangle 
 -
(\xi - i s\eta )^2 \; \langle b^{\dag}_{k\sigma s} d_{\sigma} \rangle ].
\end{equation}
We investigate the current in the adiabatic limit that
the frequency $\Omega$ of external  time-dependent sources is
much smaller than the dot-lead coupling 
$\Gamma(t)=\pi \rho |V(t)|^2$,
where $\rho$ is the density of states at
the Fermi energy in the leads:
$\Omega  \ll \Gamma(t)/\hbar$.
This condition is consistent with the one discussed in Ref.~\cite{Moskalets02}.
In this limit, Eq.~(\ref{eq:current_definition}) 
can be written as $I_{\sigma} = I_{\sigma;\rm B} + I_{\sigma;\rm F}$ with
\begin{equation}\label{eq:current_Berry}
\begin{split}
I_{\sigma; \rm B}(t) = & -\frac{2 e \Gamma(t)  \xi(t) \eta(t)}{2\pi} a(t) \\ & \times \int d \epsilon \; [G^R_{\sigma}(\epsilon;t) +
G^A_{\sigma}(\epsilon;t)]
\left( -\frac{\partial f}{\partial \epsilon} \right),
\end{split}
\end{equation}
and
\begin{equation}\label{eq:current_Friedel}
I_{\sigma;\rm F}(t)= e \frac{(\xi^2(t) - \eta^2(t))}{2} \frac{d n_{\sigma; \rm
dot}}{dt},
\end{equation}
where
 $G^{R/A}(\epsilon, t) = 1/[\epsilon - E_{0} \pm i \Gamma(t)]$ 
is the retarded (advanced) Green functions of electrons in the dot,
$f$ is the Fermi-Dirac function, and $n_{\sigma; \rm dot}$ is the occupation number in the dot.
(The derivation will be discussed later.)
We emphasize that
$I_{\sigma, B}$ is proportional to the real part of the dot Green function $G^{R}$;
in contrast 
the conductance given by the Landauer formula
is proportional to the imaginary part.
This difference indicates that the adiabatic pumping will convey
additional information on the electronic state in the dot
which is not captured by  the conductance.
We further notice that 
$I_{\sigma; F}$ always accompanies the change of 
$n_{\sigma; \rm dot}$ when it flows
while $I_{\sigma,B}$ does not.
Note  that similar expressions are obtained in a recent paper~\cite{Kashcheyevs03}
using a different formalism~\cite{Entin02_2}.
For simplicity,
we consider zero temperature limit in the following.

\begin{figure}
\includegraphics[width=7cm]{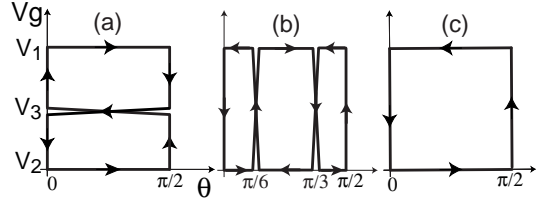}
\caption{%
Pumping  paths
in the ($\theta$, $V_{g}$) plane,
where $V_{g}$ is the gate voltage and $\theta$ defines that
$V_{\rm L}= V \cos \theta$ and
$V_{\rm R}= V \sin \theta$ with a fixed 
$V=\sqrt{V_{\rm L}^{2}+V_{\rm R}^{2}}$.
$V_{3}$ is defined through a relation for the Friedel phase $\delta$:
$\delta(V_{3})=[\delta(V_{2})+\delta(V_{1})]/2$.
}
\label{fig:path}
\end{figure}

Equations.~(\ref{eq:current_Berry}) and (\ref{eq:current_Friedel}) are
also derived by the conventional scattering matrix approach of the pumping~\cite{Brouwer98,Shutenko00,Buttiker94}.
The pumped charge $Q_{\sigma}$ per cycle is
\begin{equation}\label{eq:current_S-matrix}
Q_{\sigma} =  \frac{e}{2 \pi} \oint dt \;
 \left( 1-T_{\sigma} \right)
\frac{d \alpha_{\sigma}}{dt}
\equiv 
\oint dt I_{\sigma; S} (t)
\end{equation}
with the transmission probability $T_{\sigma}$ and
the phase of the reflection coefficient $\alpha_{\sigma}$ through the dot.
The scattering matrix of the dot is given by
\begin{equation}\label{eq:S-matrix}
U
\begin{pmatrix}
           e^{2 i \delta_{\sigma}} & 0\\
           0  & 0
\end{pmatrix}
U^{-1}=
\begin{pmatrix}
  \xi^2 e^{2 i \delta_{\sigma}} + \eta^2 &
  \xi \eta (e^{2 i \delta_{\sigma}}-1)\\
  \xi \eta (e^{2 i \delta_{\sigma}}-1)  &
  \xi^2  + \eta^2 e^{2 i \delta_{\sigma}}
\end{pmatrix}
\end{equation}
with the Friedel phase $\delta_{\sigma}$ at the Fermi energy in the leads.
It determines
$T_{\sigma} = 4  \xi^2 \eta^2 \sin \delta_{\sigma}^2$,
and
$\alpha_{\sigma} = \arctan [ (\xi^2-\eta^2) \tan \delta_{\sigma}]$.
Hence
\begin{equation}\label{eq:pumping_current_Smat}
I_{\sigma;S} = 
\frac{e}{2 \pi} 2 \xi \eta\;  a \sin 2 \delta_{\sigma} 
+ \frac{e}{2 \pi} ( \xi^2 - \eta^2) \frac{ d \delta_{\sigma}}{d t}.
\end{equation}
The first term corresponds to
Eq.~(\ref{eq:current_Berry}) and
the second to Eq.~(\ref{eq:current_Friedel}).
These are due to
(i) the relation of
$\exp(2 i \delta_{\sigma}) = 1 - 2 i \Gamma G^{R}(\epsilon=0;t)$,
and (ii) the Friedel sum rule~\cite{Langreth66}:
$n_{\sigma; \rm dot} = \delta_{\sigma}/\pi$
(See the discussion later).
Pumped charge $Q_{\sigma}$ is  then represented by
two parameters,
$\delta_{\sigma}$ and $\theta$ which defines
$\xi=\cos \theta$ and $\eta = \sin \theta$
($0 \leq \theta \leq \pi/2$):
\begin{equation}
  \label{eq:pumped_charge}
  Q_{\sigma}= \frac{e}{2\pi} \oint 
  \left( 
    \cos 2 \theta d \delta_{\sigma} - 
    \sin 2 \theta \sin 2 \delta_{\sigma} d \theta 
  \right).
\end{equation}
The Friedel phase $\delta_{\sigma}$ will be controlled
by the gate voltage $V_{g}=E_{0}-\mu$.
Furthermore, $\delta_{\sigma}$ is independent of $\theta$ since
the effect of $\theta$ is only involved in the hamiltonian~(\ref{eq:Hamiltoninan_Berry}) 
as the factor $a(t) = - d \theta/dt$,
which  is small in the adiabatic limit and can be disregarded.
Thereby,
$\delta_{\sigma}=\delta_{\sigma}(V_g)$.
In the following, we choose
$V_{g}$ and $\theta$ as the pumping parameters.
(See Fig.~\ref{fig:path}.)

{\it Pumping under the Coulomb interaction.---}
Now we investigate the effect of $u$ in the mixed-valence and Kondo regimes.
To this end,
we adopt the slave boson mean field approximation~\cite{Read83,Coleman87},
which has been successfully applied for the regimes.
We assume $u \rightarrow \infty$ to
exclude the double occupancy of electrons in the dot.
In this situation,
the annihilation operator $d_{\sigma}$ of electron is the dot
is written as $d_{\sigma} = b^{\dag} f_{\sigma}$
with
the slave boson operator $b$ and the pseudo fermion operator $f_{\sigma}$
with the constraint term of $H_c=
\lambda(\sum_{\sigma}f^{\dagger}_
{\sigma}f_{\sigma}+b^{\dagger}b-1),
$
where $\lambda$ is a Lagrange multiplier.
We assume
that $b$ and $\lambda$ are constant, determined by
the self-consistent equations as 
a function of the gate voltage $V_{g}$~\cite{Read83,Coleman87}.
After the approximation,
the dot level and dot-lead coupling are renormalized:
$\widetilde{E}_{0}= E_{0}+ \lambda$,
and
$\widetilde{\Gamma} = \Gamma b^{2}$
~\cite{Remark:SBMF_and_Kondo}.
The current is therefore given by the sum of
Eqs.~(\ref{eq:current_Berry}) and (\ref{eq:current_Friedel})
with these renormalized values.
Note that to keep the adiabatic condition,
we need a smaller value of  $\Omega$;
$\Omega \ll  \widetilde{\Gamma}/\hbar$
since  $\widetilde{\Gamma}<  \Gamma$.

\begin{figure}
\includegraphics[width=6cm]{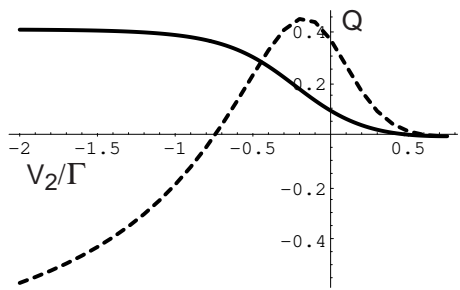}
\caption{%
Pumped charge $Q$ (in the unit of $e/2\pi$) for
the path in Fig.~\ref{fig:path}(a) as a function of $V_2$ for $V_1 = \Gamma$.
The solid and broken lines stand for
the interacting ($u\rightarrow \infty$) and non-interacting ($u=0$) models,
respectively.
}
\label{fig:sine}
\end{figure}

Let us show that 
$I_{\sigma, B}$ can flow in the opposite direction depending on
the strength of $u$.
To this end,
we choose the pumping path as shown in
Fig.~\ref{fig:path}(a),
where $Q=Q_{\sigma}=
 \frac{e}{2 \pi} \sin^2 \frac{\delta_1-\delta_2}{2} 
\sin ( \delta_1 +\delta_2)$.
(The contribution from $I_{\sigma, F}$ is cancelled.)
In Fig.~\ref{fig:sine},
$Q$ (in the unit of $e/2\pi$) is plotted as a function of
$V_2$ for $V_1=\Gamma$ ($\delta_1 < \delta < \delta_2$). 
The solid and broken lines represent the result to
for the interacting ($u \rightarrow \infty$) 
and non-interacting ($u=0$) models, respectively.
The qualitative difference  in $Q$ between the two models appears when
the electron interaction is essential ($V_{2} < -\Gamma$).
For the non-interacting model,
$Q$ is negative,
while for the interacting model,
it is positive;
the pumped current flows in the opposite directions.
If we choose the path as in Fig.~\ref{fig:path}(b),
only the contribution from $I_{\sigma; F}$ remains and
$Q =\frac{e}{\pi} (\delta_{2}-\delta_{1})$.
Then the sign change of $Q$ does not happen. 
The conductance also does not show
such a change of sign.

\begin{figure}
\includegraphics[width=6cm]{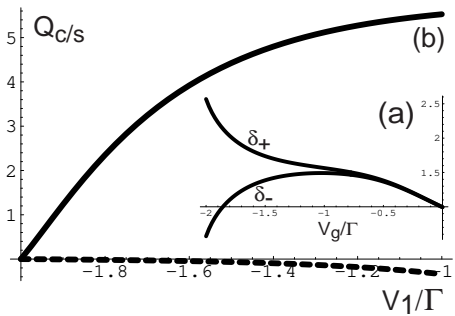}
\caption{
Electron pumping  under the Zeeman effect;
$E_{Z}=5.0\times10^{-3} \Gamma$.
(a) $\delta_{\sigma}$ as a function of the gate voltage $V_g$.
(b) The pumped charge $Q_{c}=Q_{+}+Q_{-}$ (the broken line), and
pumped spin $Q_{s}= Q_{+}-Q_{-}$ (the solid line) 
(in the unit of $e/2\pi$) for the path in Fig.~\ref{fig:path}(c) as a function of
$V_1$ for a fixed $V_2=-2.0\Gamma$.
}
\label{fig:spin_pumping}
\end{figure}

{\it Spin pumping in the Kondo regime.---}
Let us look at the Kondo effect under
a finite magnetic field,
where
the Zeeman energy $E_{Z}$ lifts
spin degeneracy of the dot level:
$E_{0} \rightarrow E_{0} \pm E_{Z}$.
In Fig.~\ref{fig:spin_pumping}(a),
the Friedel phases $\delta_{\pm}$ are plotted as a function of
the gate voltage $V_{g}$ under $E_{Z}=5.0\times10^{-3} \Gamma$.
When $V_{g} < -\Gamma$,
the Zeeman effect competes with
the Kondo effect, and
\begin{equation}\label{eq:Fridel_phase_Kondo}
 \delta_{\pm} \sim  \pi/2 \pm \Delta \delta,
\end{equation}
with a certain phase $\Delta \delta$.
This phase shift is peculiar to
the Zeeman splitting of the Kondo state;
the center of the splitting peaks is fixed at the Fermi level in the leads
~\cite{Remark:Low_gate_voltage}.
Pumped charge $Q_{\sigma}$ depends on
the spin $\sigma$ accordingly.

Now we investigate the pumping in the Kondo regime 
along the path shown in Fig.~\ref{fig:path}(c),
where $Q_{\sigma}$ is given by
\begin{equation}\label{eq:charge_simple}
  Q_{\sigma} = 
  \frac{e}{2\pi} \left[ 2 (\delta_{2,\sigma} - \delta_{1,\sigma}) -
  (  \sin 2 \delta_{2,\sigma}- \sin 2 \delta_{1,\sigma} ) \right]
\end{equation}
with $\delta_{j,\sigma} = \delta_{\sigma}(V_j) \; (j=1,2)$.
In Fig.~\ref{fig:spin_pumping}(b),
the pumped charge $Q_{c}=Q_{+}+Q_{-}$ (the broken line)
 and pumped spin 
$Q_{s}=Q_+-Q_-$ (the solid line) are plotted as
a function of $V_1$ for a fixed $V_2 = -2 \Gamma$
($ -2 \Gamma < V_1 < - \Gamma$).
We obtain $Q_c \sim 0$ and $Q_s \neq 0$;
the spin pumping without the charge pumping
~\cite{Watson03,Sharma01,Mucciolo02,Aono03,Governale03,Wang03,Zhou03,Sharma03} is realized.
This result is explained by
Eqs.~(\ref{eq:Fridel_phase_Kondo}) and (\ref{eq:charge_simple});
The effect of $\Delta \delta$ is cancelled for $Q_{c}$ and doubled for
$Q_{s}$.

The absence of charge pumping is a generic feature in the Kondo regime, and
in contrast to the conductance, which
is the maximum of $2 e^2/h$.
On the other hand,
the finite pumped spin proves that the spin degree of freedom is active.
These results are consistent with the fact that
in the Kondo regime,
the charge excitations freeze while
the spin excitations are active.
Thereby the pumping seizes this separation peculiar to the Kondo effect.

\begin{figure}
\includegraphics[width=6cm]{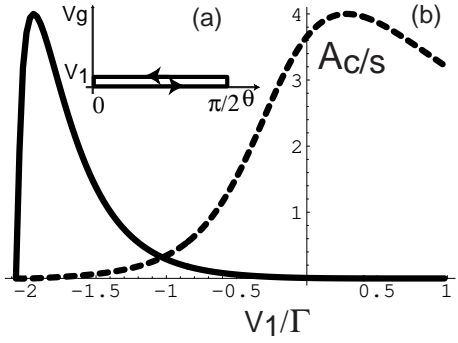}
\caption{%
(a)
A pumping path in the ($\theta,V_{g}$) plane.
(b)
The amplitude of the charge and spin alternating currents,
$A_{c}=A_{+}+A_{-}$ (the broken line), and $A_{s}=A_{+}- A_{-}$ (the solid line) as a function
of the gate voltage $V_{1}$,
where $A_{\sigma} \equiv 2 \sin (2 \delta_{\sigma})$.
}
\label{fig:ac_pumping}
\end{figure}

We further demonstrate this separation in the pumping,
investigating the pumping  along the path shown
in Fig.~\ref{fig:ac_pumping}(a);
tunneling barriers are oscillating alternatively.
This pumping path produces zero direct current but
finite alternating current.
The amplitude of this alternating current is
proportional to $ A_{\sigma} \equiv 2 \sin (2 \delta_{\sigma})$
since $\frac{d \delta_{\sigma}}{d t}=0$
[ See Eq. (\ref{eq:pumping_current_Smat}).]
In Fig.\ref{fig:ac_pumping}(b),
the amplitudes of the charge and  spin current,
$A_{c}=A_{+}+A_{-}$ (the broken line)  and
$A_{s}=A_{+}-A_{-}$ (the solid line),
are plotted as a function of $V_{1}$.
Around $V_{1} \sim 0$,
$A_{c}$ has a peak and
decreases to zero as $V_{1}$ decreases.
On the other hand,
$A_{s}$ is zero for high gate voltage and
increases as $V_{1}$ decreases, and
it has a peak around $V_{1} \sim - 2 \Gamma$.

The separation of the peak structures between $A_{c}$ and $A_{s}$ is 
a direct evidence of the separation of the spin and charge excitations
peculiar to the Kondo effect.
The peak of $A_{s}$ appears in the Kondo regime described above.
On the other hand,
the peak of $A_{c}$ appears in the mixed-valence regime,
where the number of electrons in the dot can fluctuate; the charge
excitations are active. In this way, the pumping reveals
the intrinsic nature of electronic states in the Kondo effect.
Note that
when $V_{1} \ll -\Gamma$,
where the spin polarized state appears,
 both $A_{s/c}$ are zero.

{\it Keldysh Green functions.---}
We now discuss the derivation of 
Eqs. (\ref{eq:current_Berry}) and (\ref{eq:current_Friedel}) .
Equation (\ref{eq:current_definition})
 is expressed in terms of the Keldysh Green functions~\cite{Jauho94}:
$I_{\sigma} = I_{\sigma;\rm B} + I_{\sigma;\rm F}$ with
\begin{eqnarray}
I_{\sigma; \rm B} &=& 
\frac{2 i e V \xi \eta }{4 \hbar} \;
V
(
 G^{R}_{\sigma} g^{<}_{\sigma A} + g^{<}_{\sigma A} G^{A}_{\sigma}
), \label{eq:current_Green_Berry}\\
I_{\sigma; \rm F} &=& 
 \frac{e V  (\xi^2-\eta^2) }{4 \hbar}\; \\ \nonumber
& & \times V
[
 G^{R}_{\sigma} g^{<}_{\sigma S} -  g^{<}_{\sigma S} G^{A}_{\sigma}
+ G^{<}_{\sigma} g^{R}_{\sigma S}- g^{A}_{\sigma S} G^{<}_{\sigma}
],
\label{eq:current_Green_Friedel}
\end{eqnarray}
where
we have introduced the Keldysh Green functions in the dot and leads,
$G^{j}_{\sigma}$ and $g^{j}_{\sigma k s}$ ($j=R,A,<$) ,
and
$ g^{j}_{\sigma S/A}= 
\sum_k (  g^{j}_{ \sigma k +} \pm g^{j}_{\sigma k -} )$.
The notation of
$V G^{R}_{\sigma} g^{<}_{\sigma A}$ stands for
$ \int dt_1
V(t_1) G^{R}_{\sigma}(t,t_1) g^{<}_{\sigma A}(t_1,t)$
for example.

To investigate adiabatic pumping regime,
we have assumed the following four points:
(a)
$
g^{<}_{\sigma k s}(t,t') = g^{<}_{\sigma k s}(t-t';t)=  i  f(\epsilon_{k} - \mu(t) -s \hbar a(t))
\exp[- i \epsilon_{k}/\hbar (t-t')]$
which means that
the leads keep thermal equilibrium with
the time-dependent chemical potential $\mu(t)\pm \hbar a(t)$.
(b)
$
G^{R/A}_{\sigma}(t,t') =G^{R/A}_{\sigma}(t-t';t)=
\mp i \;\theta(\pm t \mp t')
\exp[- [iE_{\sigma} \pm \Gamma(t)]/ \hbar (t-t')]$,
the Fourier transformation of which is given just below
Eq.~\ref{eq:current_Friedel}.
(c)
$V(t) V(t') \simeq V(t)^{2}$ under the integration.
(d) 
$
f(\epsilon - \mu(t) - s \hbar a(t)) \simeq f(\epsilon - \mu(t)) -
\frac{\partial f(\epsilon - \mu(t))}{\partial \epsilon} s \hbar a(t).
$
The first three  assumptions rely on the relation of
$\Omega  \ll \Gamma(t)/ \hbar$ and
the system follows instantaneous values of $V_{L/R}$ and $\mu$.

The above assumptions simplify the expressions of $I_{\sigma;B}$ 
as in Eq.~(\ref{eq:current_Berry})~\cite{Remark: Keldysh}.
By the assumption (d),
$I_{\sigma;\rm F}$ is independent of $a(t)$  and accordingly
the resulting expression represents the current
between the dot and double leads that have a common chemical potential $\mu(t)$.
Then we can disregard the current between the two leads in the expression,
and it is always detected by the change of $n_{\sigma; \rm dot}$.
In general, the first two terms in Eq.~(\ref{eq:current_Green_Friedel})
represent the current
from the leads to the dot while 
the rest two terms the current  from the dot to leads~\cite{Jauho94}.
The sum of two terms therefore yields
the time derivative of $n_{\sigma; \rm dot}$.
Note that 
we can finally apply the Friedel sum rule~\cite{Langreth66} to
relate  $n_{\sigma; \rm dot}$ with $\delta_{\sigma}$ 
at this point.

In conclusion,
we have investigated
adiabatic electron pumping
with time dependent barriers under the electron interactions and
shown the separation of the pumped charge and spin 
peculiar to the Kondo effect.
The adiabatic pumping will be  a new probe to
disclose the electron interactions,
which is not achieved by
conventional conductance measurements.

It is a pleasure to acknowledge discussions
with Y.~Avishai, A.~Golub, V.~Kashcheyevs, and especially
D.~Cohen.
This work is supported by
the JSPS  Postdoctoral Fellowships.


\begin{thebibliography}{99}
\bibitem{Remark:SBMF_and_Kondo}
When the system is in the Kondo regime,
$\widetilde{E}_{0}$ gives the position of 
the Kondo resonance,
$\widetilde{E}_{0} \simeq 0$, and
$\widetilde{\Gamma}$ gives
the Kondo temperature 
$T_{K}= \frac{\Delta}{\pi} \exp ( \pi V_{g}/\Delta)$.
\bibitem{Remark:Low_gate_voltage}
When $V_{g}  \ll -\Gamma$,
the system in
the spin-polarized state rather than the Kondo state,
and as a result,
$\delta_{+}=\pi$ and $\delta_{-}=0$.
[This regime is not included in the Fig.~\ref{fig:spin_pumping}(a).]
\bibitem{Remark: Keldysh}
The discrete sum of $k$ in $\sum_{k}$ can be expressed
as $ \rho \int d \epsilon$.
We have also used the relations of
$g^{R/A}_{\sigma k s}(\omega) =
\mp i \;\theta(\pm t \mp t')
\exp[- i \epsilon_{k} (t-t')]$.
\end{thebibliography}
\end{document}